\documentclass[11pt,twoside,psfig]{article}
\usepackage{asp2006}
\usepackage{psfig}
\usepackage{lscape}
\usepackage{graphicx}

\def\NH     {\hbox{$N_{\rm H}$}}

\def    \beq       {\begin{equation}}
\def    \eeq       {\end{equation}}

\def    \Angstrom  {\,{\rm\AA}}

\def    \K      {\,{\rm K}}

\def    \NH     {N_{\rm H}}

\def	\mum	{\,\mu{\rm m}}
\def	\um	{\,\mu{\rm m}}
\def	\simali	{\sim\,}

\def	\ppm    {\,{\rm ppm}}

\def    \NH     {N_{\rm H}}

\def    \magni  {\,{\rm mag}}
\def    \simlt  {\lesssim}      
\def    \simgt  {\gtrsim}       

\def\lesssim{\mathrel{\hbox{\rlap{\hbox{\lower4pt\hbox{$\sim$}}}\hbox{$<$}}}}
\def\gtrsim{\mathrel{\hbox{\rlap{\hbox{\lower4pt\hbox{$\sim$}}}\hbox{$>$}}}}

\makeatletter
\def\@jourvol{373}
\def\cpr@year{Xi'an, China, 16--21 October, 2006}
\def\vol@title{The Central Engine of Active Galactic Nuclei}
\def\vol@author{eds. L. C. Ho and J.-M. Wang}
\makeatother

\begin{document}

\title{Dust in Active Galactic Nuclei}
\vspace{-5mm}
\author{Aigen Li}
\affil{Department of Physics and Astronomy,
       University of Missouri,
       Columbia, MO 65211, USA; 
       email: {\sf lia@missouri.edu}}

\begin{abstract}
Dust plays an essential role in the unification theory
of active galactic nuclei (AGNs).
This review summarizes our current understanding
of the extinction and infrared emission properties 
of the circumnuclear dust in AGNs 
as well as the inferred dust composition 
and size distribution.
\end{abstract}

\vspace{-9mm}
\section{Introduction: Are All AGNs Born Equal?
--- The Role of Dust in the Unified Schemes of AGNs}
\vspace{-3mm}
Dust is the cornerstone of the unification theory
of active galactic nuclei (AGNs). This theory proposes
that all AGNs are essentially ``born equal'':
all types of AGNs are surrounded by an optically thick
dust torus and are basically the same object
but viewed from different lines of sight
(see e.g. Antonucci 1993; Urry \& Padovani 1995).
The large diversity in the observational properties 
of AGNs (e.g. optical emission-line widths 
and X-ray spectral slopes) is simply caused by 
the viewing-angle-dependent obscuration of the nucleus: 
those viewed face-on are unobscured (allowing for 
a direct view of their nuclei) and recognized as 
``type 1'' AGNs, while those viewed edge-on are 
``type 2'' AGNs with most of their central engine and
broad line regions being hidden by the obscuring dust.

Apparently, key factors in understanding the structure 
and nature of AGNs are determining the geometry of 
the nuclear obscuring torus around the central engine 
and the obscuration (i.e. extinction, a combination 
of absorption and scattering) properties of 
the circumnuclear dust.
An accurate knowledge of the dust extinction properties
is also required to correct for the dust obscuration
in order to recover the intrinsic optical/ultraviolet (UV)
spectrum of the nucleus from the observed spectrum 
and to probe the physical conditions of 
the dust-enshrouded gas close to the nucleus.

The presence of an obscuring dust torus around
the central engine was first indirectly indicated
by the spectropolarimetric detection of 
broad permitted emission lines (characteristic of 
type 1 AGNs) scattered into our line of sight by free electrons
located above or below the dust torus 
in a number of type 2 AGNs 
(e.g. see Heisler et al.\ 1997, Tran 2003). 
%
Direct evidence for the presence of a dust torus
is provided by infrared (IR) observations.
The circumnuclear dust absorbs the AGN illumination
and reradiates the absorbed energy in the IR.
The IR emission at wavelengths longward of 
$\lambda$\,$>$\,1$\mum$ accounts for at least 
50\% of the  bolometric luminosity of type 2 AGNs.
For type 1 AGNs, $\simali$10\% of the bolometric 
luminosity is emitted in the IR
(e.g. see Fig.\,13.7 of Osterbrock \& Ferland 2006).
A near-IR ``bump'' (excess emission above 
the $\simali$2--10$\mum$ continuum),
generally attributed to hot dust
with temperatures around $\simali$1200--1500$\K$
(near the sublimation temperatures of silicate
and graphite grains), is seen in a few type 1 AGNs
(Barvainis 1987; Rodr\'{i}guez-Ardila \& Mazzalay 2006).
Direct imaging at near- and mid-IR wavelengths has
been performed for several AGNs and provides constraints
on the size and structure of the circumnuclear dust torus
(e.g. see Jaffe et al.\ 2004, Elitzur 2006). 
Spectroscopically,
the 10$\mum$ silicate {\it absorption} feature (see \S3.3)
and the 3.4$\mum$ aliphatic hydrocarbon {\it absorption}
feature (see \S3.2) are widely seen in heavily obscured type 2 AGNs;
in contrast, the 10$\mum$ silicate {\it emission} feature 
has recently been detected in a number of type 1 AGNs (see \S3.3).

To properly interpret the observed IR continuum emission
and spectroscopy as well as the IR images of AGNs, it requires
a good understanding of the absorption and emission properties
of the circumnuclear dust. To this end, one needs to know
the composition, size, and morphology of the dust 
-- with this knowledge,
one can use Mie theory (for spherical dust) to calculate 
the absorption and scattering cross sections of the dust
from X-ray to far-IR wavelengths,
and then calculate its UV/optical/near-IR obscuration as 
a function of wavelength, and derive the dust thermal 
equilibrium temperature (based on the energy balance between 
absorption and emission) as well as its IR emission spectrum. 
This will allow us to correct for dust obscuration
and constrain the circumnuclear structure through 
modeling the observed IR emission and images. 
The former is essential for interpreting the obscured
UV/optical emission lines and probing the physical 
conditions of the central regions;
the latter is critical to our understanding 
of the growth of the central supermassive black hole.

However, little is known about the dust in the circumnuclear
torus of AGNs. Even our knowledge of the best-studied dust 
-- the Milky Way interstellar dust -- is very limited.
In this review, I will take a comparative study of 
the extinction and IR emission 
as well as the UV/IR spectroscopic properties 
and the inferred composition, size and morphology 
of the dust in AGNs and the dust in the interstellar
medium (ISM) of the Milky Way and other galaxies.

\vspace{-4.5mm}
\section{Extinction --- A Powerful Discriminator of Dust Size}
\vspace{-3mm}
Extinction is a combined effect of absorption and scattering.
Since a grain absorbs and scatters light most effectively 
at wavelengths comparable to its size 
$\lambda$$\approx$$2\pi a$, 
the wavelength dependence of extinction 
(``extinction curve'') constrains the dust size
distribution.

\vspace{-2.5mm}
\subsection{Interstellar Extinction: Milky Way, SMC, and LMC}
\vspace{-1mm}
Interstellar extinction is most commonly 
obtained through the ``pair-method'' by comparing 
the spectra of two stars of the same spectral type, 
one of which is reddened and the other unreddened.  
Interstellar extinction curves rise from 
the near-IR to the near-{\small UV}, 
with a broad absorption feature at about 
$\lambda^{-1}$$\approx$4.6$\mum^{-1}$ 
($\lambda$$\approx$2175$\Angstrom$),
followed by a steep rise into the far-{\small UV}
$\lambda^{-1}$$\approx$10$\mum^{-1}$ (see Fig.\,1). 
This wavelength dependence indicates that
there must exist in the ISM a population 
of large grains with $a$$\simgt$$\lambda/2\pi$$\approx$$0.1\mum$
to account for the extinction at visible/near-IR wavelengths,
and a population of ultrasmall grains 
with $a$$\simlt$$\lambda/2\pi$$\approx$\\$0.016\mum$
to account for the far-{\small UV} extinction at 
$\lambda$\,=\,$0.1\mum$. 
In the wavelength range of 0.125\,$\le$$\lambda$$\le$\,3.5$\mum$,
the Galactic extinction curves can be approximated by 
an analytical formula involving 
only one free parameter: {\small $R_V$$\equiv$$A_V/E(B-V)$}, 
the total-to-selective extinction ratio
(Cardelli et al.\ 1989),
with $R_V$\,$\approx$\,3.1 
for the Galactic average (see Fig.\,1). 
The optical/{\small UV} extinction curves and $R_V$ show
considerable regional variations and 
depend on the environment: lower-density regions have 
a smaller {\small $R_V$}, 
a stronger {\small 2175$\Angstrom$} bump 
and a steeper far-{\small UV} rise 
($\lambda^{-1}$\,$>$\,4$\mum^{-1}$),
implying smaller dust in these regions;
denser regions have a larger {\small $R_V$}, 
a weaker {\small 2175$\Angstrom$} bump 
and a flatter far-{\small UV} rise, implying larger dust.

\begin{figure} 
\vspace*{-1.0em}
\centerline{
\psfig{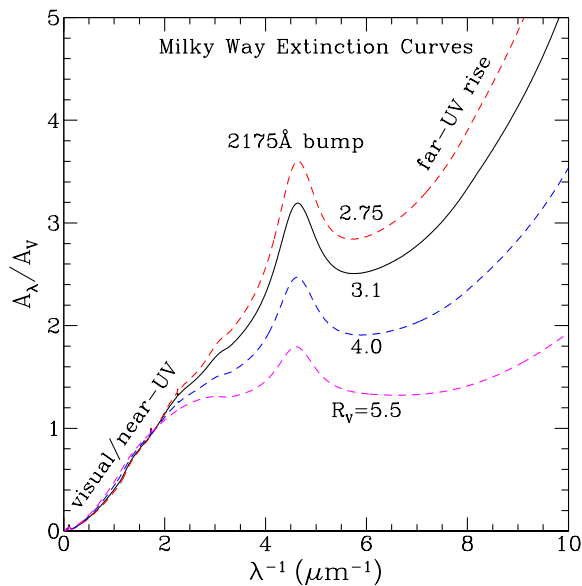}}
\vspace*{-1.6em}
\caption{
        \normalsize
        Interstellar extinction curves of
        the Milky Way ($R_V$\,=\,2.75, 3.1, 4.0, 5.5).
        There exist considerable regional variations 
        in the Galactic optical/UV extinction curves,
        as characterized by the total-to-selective 
        extinction ratio $R_V$, indicating that
        dust grains on different sightlines have 
        different size distributions. 
        }
\vspace*{-1.3em}
\end{figure}

In the Small Magellanic Cloud ({\small SMC}), 
the extinction curves of most sightlines display 
a nearly linear steep rise with $\lambda^{-1}$ 
and an extremely weak or absent 2175$\Angstrom$ bump 
(Lequeux et al.\ 1982; Pr\'evot et al.\ 1984; see Fig.\,2),
suggesting that the dust in the SMC is smaller than
that in the Galactic diffuse ISM as a result of either
more efficient dust destruction in the {\small SMC} 
due to its harsh environment of the copious star formation 
associated with the {\small SMC} Bar 
or lack of growth due to 
the low-metallicity of the {\small SMC}, or both.
The Large Magellanic Cloud ({\small LMC}) extinction curve 
is characterized by a weaker 2175$\Angstrom$ bump
and a stronger far-{\small UV} rise than 
the Galactic curve
(Nandy et al.\ 1981; Koornneef \& Code 1981), 
intermediate between that of the {\small SMC} 
and that of the Galaxy (see Fig.\,2). 
Regional variations also exist 
in the SMC and LMC extinction curves.

\begin{figure} 
\vspace*{-1.2em}
\centerline{
\psfig{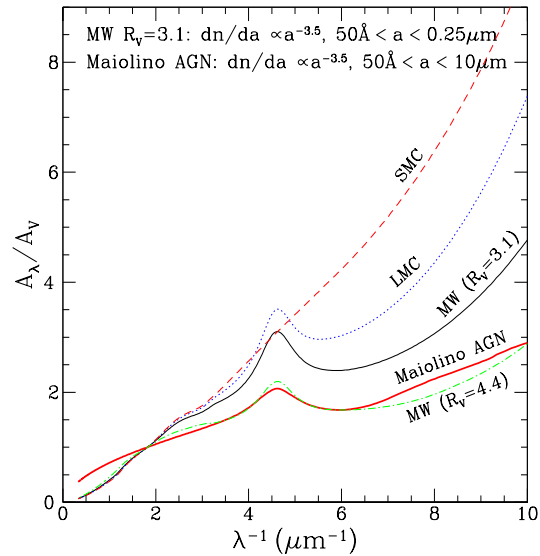}}
\vspace*{-1.7em}
\caption{
        \normalsize
        Interstellar extinction curves of
        the Milky Way ($R_V$\,=\,3.1, 4.4),
        SMC, and LMC. Also plotted is the ``Maiolino''-type
        extinction curve for AGNs (similar to the Galactic 
        $R_V$\,=\,4.4 curve) produced by 
        a mixture of interstellar silicate and graphite grains
        with a size distribution of $dn/da\sim a^{-3.5}$
        (50$\Angstrom$\,$<$\,$a$\,$<$\,10$\mum$),
        the same as that for the Galactic average $R_V$\,=\,3.1
        except with a smaller upper cutoff for the latter
        (50$\Angstrom$\,$<$\,$a$\,$<$\,0.25$\mum$).
        }
\vspace*{-2.4em}
\end{figure}

\vspace{-2mm}
\subsection{AGN Extinction --- ``Gray'' or SMC-like Extinction?}
\vspace{-2mm}
Little is known about the wavelength dependence of 
the extinction caused by the circumnuclear dust of AGNs.
In literature, the AGN extinction curves are mainly
inferred from (1) composite quasar spectra,
and (2) individual reddened AGNs.
The former often reveals a ``gray'' extinction, 
implying that the size distribution
of the dust in the AGN circumnuclear environments  
is skewed towards substantially large grains.
The latter often suggests a steep-rising SMC-like extinction,
indicating a preponderance of small grains near the nucleus. 
There is also indirect information, 
including the dust reddening-
and extinction-to-gas ratios 
and the IR emission modeling of AGNs (see \S4).

\vspace{-1mm}
\subsubsection{Composite Reddened Quasar Spectra --- ``Gray'' Extinction?}
\vspace{-2mm}
Czerny et al.\ (2004) constructed a quasar extinction curve
based on the blue and red composite quasar spectra of
Richards et al.\ (2003) obtained 
from the Sloan Digital Sky Survey (SDSS).
Six composite quasar spectra
were generated by Richards et al.\ (2003)
from 4576 SDSS quasars 
based on the relative g$^{\ast}$--i$^{\ast}$ color
with ``Composite 1'' (made from 770 objects) being the bluest.
Czerny et al.\ (2004) created a mean ``quasar extinction curve''
by averaging 3 extinction curves obtained   
through comparing the spectra of Composites 3, 4, and 5 
(consisting of 770, 770, and 211 objects, respectively)
with that of Composite 1, assuming that Composite 1
is essentially unaffected by dust 
while Composites 3, 4, and 5 are subject to dust reddening.
The resulting extinction curve is nearly monotonic 
with wavelength, without any trace of 
the 2175$\Angstrom$ bump (see Fig.\,3). 

Gaskell et al.\ (2004) derived extinction curves 
for radio-loud quasars based on the composite spectra
of 72 radio quasars created by Baker \& Hunstead (1995),
and for radio-quiet AGNs based on the composite spectrum
of 1018 radio-quiet AGNs generated by Francis et al.\ (1991).
The extinction curve for these radio-loud quasars,
grouped by Baker \& Hunstead (1995) into 4 subsamples 
according to the 5\,GHz radio core-to-lobe flux ratios $R$,
was determined by comparing the composite spectrum of 
the more-reddened lobe-dominant ($R$\,$<$0.1) sample
with that of less-reddened core-dominant ($R$\,$>$1) sample.
Similarly, Gaskell et al.\ (2004) obtained an extinction curve 
for radio-quiet AGNs by comparing the composite spectrum 
of Francis et al.\ (1991) created for 1018 radio-quiet AGNs 
with that for the relatively unreddened
core-dominant composite of Baker \& Hunstead (1995). 
Most prominently, the derived extinction curves 
for both radio-loud quasars and radio-quiet quasarssy
lack the 2175$\Angstrom$ bump and are essentially ``gray'', 
i.e., significantly flatter in the UV than that of 
the Milky Way diffuse ISM, although it appears that 
for the latter the reddening curve 
is slightly steeper in the UV (see Fig.\,3).

However, Willott (2005) questioned the validity 
of the approach based on the ratios
of reddened and unreddened composite quasars
(Czerny et al.\ 2004; Gaskell et al.\ 2004)
since composite spectra combine quasars at different redshifts,
while the quasars going into a composite spectrum 
have a negative correlation between reddening and redshift, 
and quasar surveys in practice contain more highly reddened 
quasars at lower redshifts.
He argued that since the quasars contributing to 
the composite in the UV have typically lower reddening 
than those contributing in the optical, 
the gray UV extinction laws derived using composite quasars 
(Czerny et al.\ 2004; Gaskell et al.\ 2004) might be artificial,
and the actual AGN extinction curve may be SMC-like.



\begin{figure} 
\vspace*{-1.4em}
\centerline{
\psfig{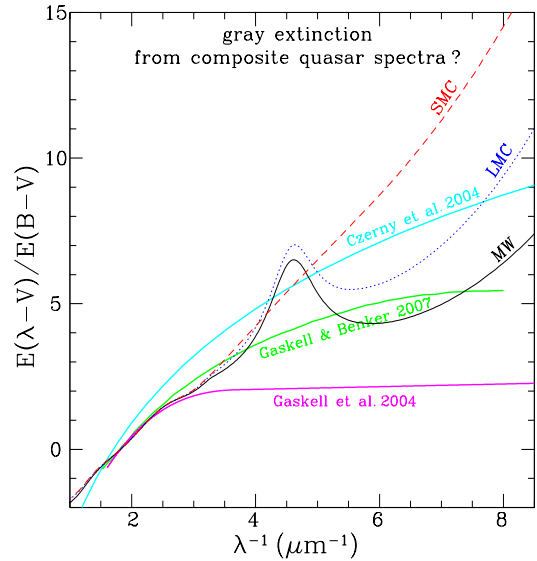}}
\vspace*{-1.7em}
\caption{
        \normalsize
        Comparison of the AGN extinction curves derived from 
        composite quasar spectra (Czerny et al.\ 2004,
        Gaskell et al.\ 2004) with that for the Milky Way,
        SMC, and LMC. For comparison, also shown is the average 
        extinction curve obtained for individual AGNs by
        Gaskell \& Benker (2007).        
        }
\vspace*{-1.2em}
\end{figure}

\vspace{-1mm}
\subsubsection{Individual Reddened AGNs --- SMC-like Extinction?}
\vspace{-2mm}
In contrast to the ``composite quasar spectrum'' method which may be 
biased by the fact that the highest redshift quasars (which contribute 
to the UV part of a composite spectrum) are less extincted 
(leading to shallower extinction in the UV), AGN extinction curves have
also been derived for individual reddened objects.

Crenshaw et al.\ (2001) determined a reddening curve for 
the nucleus of the Seyfert 1 galaxy NGC\,3227 by comparing 
its HST/STIS UV and optical spectra with that of the unreddened 
Seyfert galaxy NGC\,4151. They found that the derived extinction 
curve in the UV is even steeper than that of the SMC and lacks 
the 2175$\Angstrom$ bump.
Similar studies were performed for Ark\,564, a Narrow-Line Seyfert 1 galaxy
(Crenshaw et al.\ 2002). By comparing the HST/STIS UV and optical spectra 
of Ark\,564 with that of Mrk\,493, an unreddened Narrow-Line Seyfert 1 galaxy,
Crenshaw et al.\ (2002) found that the extinction curve for Ark\,564,
with no evidence for the 2175$\Angstrom$ bump, 
rises to the UV more steeply than the Galactic extinction curve 
(but not as steeply as the SMC curve) 
with a longer turning-up wavelength 
of $\simali$4000$\Angstrom$
(compared to $\simali$2500$\Angstrom$ 
for the standard Galactic, LMC, and SMC curves). 

In an analysis of the optical/UV color distribution of 4576 SDSS quasars,
Richards et al.\ (2003) showed that 273 (6.0\%) of the quasars in their
sample appear to be redder because of SMC-like dust extinction and reddening.
Hopkins et al.\ (2004) investigated the reddening law toward 9566 SDSS quasars,
including a subset of 1886 quasars matched to 2MASS (Two Micron All Sky Survey)
by exploring the shapes of their spectral energy distributions obtained
from broadband photometry 
(at five SDSS bands $ugriz$ and three 2MASS bands $JHK$). 
They found that the reddening toward quasars is dominated by 
SMC-like dust at the quasar redshift.

More recently, Gaskell \& Benker (2007) determined the extinction curves for
14 individual AGNs based on the FUSE and HST spectrophotometry of 
Shang et al.\ (2005). Unlike Crenshaw et al.\ (2001, 2002) who used
a single unreddened AGN as a reference, Gaskell \& Benker (2007)
took the average of 3 AGNs which have the highest 4--8$\mum^{-1}$ fluxes
relative to their optical fluxes in the sample of Shang et al.\ (2005). 
They found that the majority of the derived extinction curves in the UV
are much flatter than that of the SMC, although not as flat as 
the ``gray'' curve derived by Gaskell et al.\ (2004) based on 
composite quasar spectra (see Fig.\,3 for the average extinction curve 
for the 5 AGNs with the greatest reddening in their sample).

\vspace{-1mm}
\subsubsection{Reduced Reddening- and Extinction-to-Gas Ratios 
               --- Flat Extinction?}
\vspace{-2mm}
Assuming a Galactic standard extinction curve ($R_V$\,=\,3.1)
and a foreground screen, 
Maiolino et al.\ (2001a) determined for 19 AGNs 
the amount of reddening $E(B-V)$ 
affecting the broad line region 
by comparing the observed optical/IR H broad line ratios 
with the intrinsic values.
For these AGNs, they also determined 
the X-ray absorbing column densities $\NH$ from 
the photoelectric cutoff in their X-ray spectra. 
They found that for most (16 of 19) objects 
$E(B-V)/\NH$ is significantly
lower than the Galactic standard value
($\approx$\,$1.7\times 10^{-22}$\,mag\,cm$^{-2}$)
by a factor ranging from a few to $\simali$100
(except for 3 Low Luminosity AGNs whose physics may 
be intrinsically different [see Ho 1999]).
Similarly, Maiolino et al.\ (2001a) also found 
that the extinction-to-gas ratios $A_V/\NH$ of 
various classes of AGNs are significantly lower 
than the Galactic standard value 
($\approx$\,$5.3\times 10^{-22}$\,mag\,cm$^{-2}$).
Maiolino et al.\ (2001b) ascribed the reduced 
$E(B-V)/\NH$ and $A_V/\NH$ ratios of AGNs 
(often with a solar or higher metallicity)
to grain growth through
coagulation in the dense circumnuclear region which 
results in a dust size distribution biased in favour
of large grains and therefore a flat extinction curve.

However, Weingartner \& Murray (2002) argued that 
the X-ray absorption and optical extinction may occur 
in distinct media (e.g. the X-ray absorption occurs 
in material located off the torus and/or accretion disk, 
while the optical extinction occurs in material 
located beyond the torus); therefore, the reduced 
$E(B-V)/\NH$ and $A_V/\NH$ ratios may not necessarily
imply that the grains in AGNs are systematically larger 
than those in the Galactic ISM.

%
%

\vspace{-5.5mm}
\section{Dust Spectroscopy  --- Diagnosis of Dust Composition}
\vspace{-3.5mm}
Dust spectroscopy provides the most diagnostic information 
on the dust composition.
Our knowledge about the composition of the dust 
in the Galactic diffuse ISM is mainly derived from
the absorption and emission spectral lines:
the 2175$\Angstrom$ extinction bump 
(small graphitic dust),
the 3.4$\mum$ absorption feature
(aliphatic hydrocarbon dust),
the 9.7$\mum$ and 18$\mum$ absorption features 
(amorphous silicate dust),
and the 3.3, 6.2, 7.7, 8.6, and 11.3$\mum$ emission features
(polycyclic aromatic hydrocarbon [PAH] molecules).
The ice absorption features at
3.1 and 6.0$\mum$ (H$_2$O),
4.67$\mum$ (CO),
4.27 and 15.2$\mum$ (CO$_2$), 
3.54 and 9.75$\mum$ (CH$_3$OH),
2.97$\mum$ (NH$_3$),
7.68$\mum$ (CH$_4$),
5.81$\mum$ (H$_2$CO),
and 4.62$\mum$ (XCN$^{-}$)
are seen in dark molecular clouds 
with visual extinction $A_V$\,$>$\,3\,mag.
In this section I will present a comparative overview of
the dust absorption and emission features 
in AGNs and the inferred dust composition.
 
\vspace{-3.5mm}
\subsection{The 2175$\Angstrom$ Extinction Bump}
\vspace{-1.5mm}
The 2175$\Angstrom$ extinction bump, first detected over 40 years ago
(Stecher 1965), is an ubiquitous feature of the Milky Way ISM.
With a stable central wavelength and variable feature strength
for lines of sight in our Galaxy, the 2175$\Angstrom$ bump is 
relatively weaker in the LMC and absent in the SMC (see Fig.\,2). 
This bump is largely absent in AGNs (see \S2)
except Gaskell \& Benker (2007) recently claimed that it might 
be detected in Mrk\,304, one of the seven AGNs with the highest
quality extinction curves in their 14-AGN sample.
Fig.\,4 compares the UV spectra of 5 slightly reddened type 1 AGNs
with the template of type 1 AGNs reddened with 
the standard Galactic extinction.
It is seen that the Galactic extinction predicts 
too strong a 2175$\Angstrom$ dip (Maiolino et al.\ 2001b). 

\vspace*{-2.0em}
\begin{figure} 
\vspace*{-1.0em}
\centerline{
\psfig{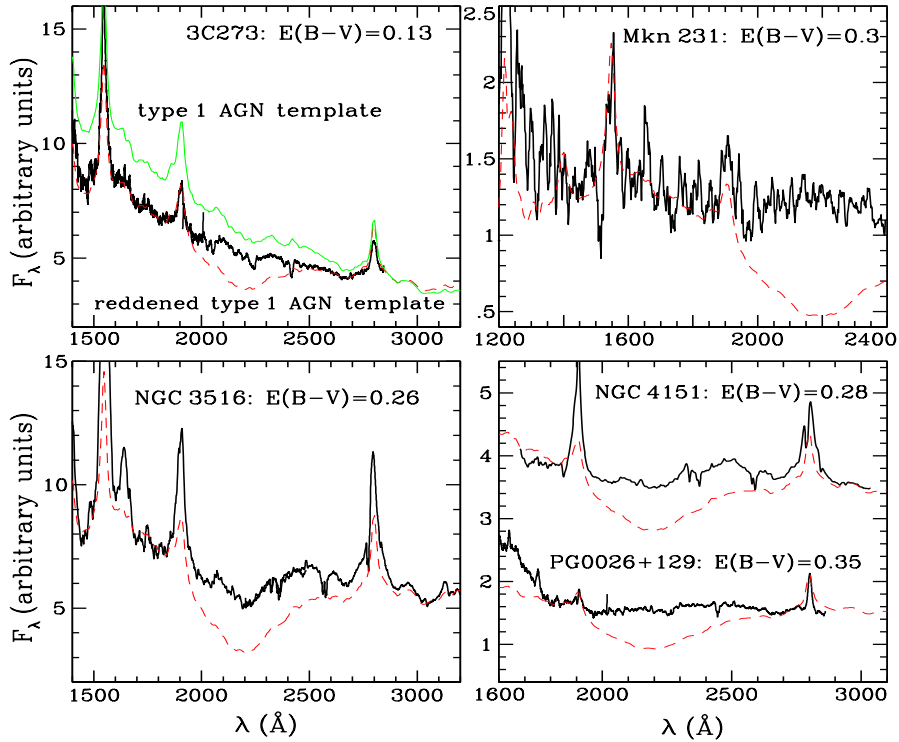}}
\vspace*{-1.2em}
\caption{
        \normalsize
        Comparison of the UV spectra (solid lines) 
        of 5 type\,1 AGNs 
        (whose broad line ratios and continuum suggest
        dust absorption) with the template of type\,1
        AGNs reddened by a standard Galactic extinction
        ($R_V$\,=\,3.1) with a total amount of reddening
        $E(B-V)$ consistent with that inferred from the
        broad lines and adapted to match the shape of
        the continuum in those regions not affected by 
        the 2175$\Angstrom$ bump (dashed line).
        Taken from Maiolino et al.\ (2001a) with 
        modifications.
        }
\vspace*{-1.2em}
\end{figure}

The exact nature of the 2175$\Angstrom$ bump,
the strongest spectroscopic extinction feature in the Galactic ISM,
remains uncertain. It is generally believed to be caused by aromatic 
carbonaceous (graphitic) materials, very likely a cosmic mixture of 
PAH molecules (Joblin et al.\ 1992; Li \& Draine 2001b).
The fact that the 2175$\Angstrom$ bump is not (or at least rarely)
seen in AGNs suggests that its carrier (e.g. PAHs) may have been
photodestroyed by energetic photons (e.g. X-ray irradiation) from
the central engine.

\vspace{-2mm}
\subsection{The 3.4$\mum$ Aliphatic Hydrocarbon Absorption Feature}
\vspace{-1mm}
The 3.4$\mum$ absorption feature, 
attributed to the C--H stretching 
mode in saturated aliphatic hydrocarbon dust,
is widely seen in the Galactic diffuse ISM
(but never seen in molecular clouds;
see Pendleton \& Allamandola 2002).
This feature is also seen in AGNs
(Wright et al.\ 1996, Imanishi et al.\ 1997, Mason et al.\ 2004),
closely resembling that of our Galaxy 
in both peak wavelengths and relative feature 
strengths of the 3.42$\um$, 3.48$\um$, 
and 3.51$\um$ subfeatures 
(corresponding to symmetric and asymmetric 
stretches of C--H bonds in CH$_2$ and CH$_3$ groups
in aliphatic hydrocarbon chains).
Mason et al.\ (2004) argued that the 3.4$\mum$ absorption
feature at least in face-on Seyfert 2 galaxies 
arises in dust local to the active nucleus
rather than in the diffuse ISM of the galaxy.

The exact carrier of this feature remains uncertain.
So far, among the $>$\,20 candidate materials proposed
over the years since its first detection in the Galactic 
center sightlines 28 years ago,
the experimental spectra of hydrogenated
amorphous carbon (Mennella et al.\ 1999)
and the organic refractory residue,
synthesized from UV photoprocessing of interstellar ice mixtures
(Greenberg et al.\ 1995), 
provide the best fit to the observed spectra.

So far, no polarization has been detected for this feature 
(Adamson et al.\ 1999, Chiar et al.\ 2006, Mason et al.\ 2006),
suggesting that the carrier of this feature is either spherical
or unaligned or both.  
Spectropolarimetric measurements for 
both the 9.7$\mum$ silicate 
and the 3.4$\mum$ hydrocarbon features 
for the same sightline 
(e.g. Chiar et al.\ 2006) would allow for a direct test of 
the silicate core-hydrocarbon mantle interstellar dust model
(Li \& Greenberg 1997, Jones et al.\ 1990), 
since this model predicts that the 3.4$\mum$ feature 
would be polarized if the 9.7$\mum$ feature 
(for the same sightline) is polarized 
(Li \& Greenberg 2002).

\vspace{-2mm}
\subsection{The 9.7$\mum$ and 18$\mum$ Silicate 
            Absorption and Emission Features}
\vspace{-1mm}
The strongest IR absorption features in the Galactic ISM 
are the 
9.7$\mum$ and 18$\mum$ bands, 
which are almost certainly due to silicate minerals: 
they are respectively ascribed to the Si--O stretching 
and O--Si--O bending modes in some form of silicate material 
(e.g. olivine Mg$_{2x}$Fe$_{2-2x}$SiO$_4$).
The observed interstellar silicate bands are broad and 
relatively featureless, indicating that interstellar 
silicates are largely amorphous rather than crystalline
(Li \& Draine [2001a] estimated that the amount of 
$a$\,$<$\,1$\mum$ crystalline silicate grains in 
the Galactic diffuse ISM is $<$\,5\% of the solar Si abundance).

The first detection of the silicate {\it absorption} 
feature in AGNs was made at 9.7$\mum$ for 
the prototypical Seyfert 2 galaxy NGC\,1068
(Rieke \& Low 1975; Kleinmann et al.\ 1976),
indicating the presence of a large column of silicate dust 
in the line-of-sight to the nucleus.
It is known now that most of the type 2 AGNs display 
silicate {\it absorption} bands 
(e.g. see Roche et al.\ 1991,
Siebenmorgen et al.\ 2004)
as expected -- for a centrally heated optically thick torus 
viewed edge-on, the silicate features should be in absorption.
Spatially resolved mid-IR spectra obtained for
NGC\,1068 (Mason et al.\ 2006, Rhee \& Larkin 2006)
and Circinus (Roche et al.\ 2006)
have revealed striking variations in continuum slope,
silicate feature profile and depth.

However, it appears that the 9.7$\mum$ silicate absorption
profile of AGNs differs from that of the Milky Way.
Jaffe et al.\ (2004) found that the 9.7$\mum$ silicate 
absorption spectrum of NGC\,1068 shows 
a relatively flat profile from 8 to 9$\mum$ 
and then a sharp drop between 9 and 10$\mum$;
in comparison, the Galactic silicate absorption profiles
begin to drop already at $\simali$8$\mum$.
They obtained a much better fit to the 9.7$\mum$ absorption
feature of NGC\,1068 by using the profile
of calcium aluminium silicate Ca$_2$Al$_2$SiO$_7$, 
a high-temperature dust species found
in some supergiant stars (Speck et al.\ 2000).
It would be interesting to know if the amount of
calcium required to account for the observed absorption
is consistent with abundance constraints. 
Very recently, Roche et al.\ (2007) reported 
the detection of a spectral structure near 11.2$\mum$
in NGC\,3094, indicative of the possible presence 
of crystalline silicates in AGNs.

For type 1 AGNs viewed face-on, one would expect to see 
the silicate features in {\it emission} since the silicate dust 
in the surface of the inner torus wall will be heated to 
temperatures of several hundred kelvin by the radiation 
from the central engine, allowing for a direct detection 
of the 9.7$\mum$ and 18$\mum$ silicate bands emitted from 
this hot dust. However, their detection (using {\it Spitzer}) 
has only very recently been reported in a number of type 1 AGNs 
(Hao et al.\ 2005, Siebenmorgen et al.\ 2005, 
Sturm et al.\ 2005, Weedman et al.\ 2005, Shi et al.\ 2006).
Siebenmorgen et al.\ (2005) postulated that the AGN luminosity
determines whether the silicate emission bands are prominent or not
(i.e., they may be present only in the most luminous AGNs),
but this idea was challenged by their detection 
in the low-luminosity AGN NGC\,3998, a type 1 LINER galaxy
(Sturm et al.\ 2005).

The 9.7$\mum$ silicate emission profiles of both quasars 
(high luminosity counterparts of Seyfert 1 galaxies; 
Hao et al.\ 2005, Siebenmorgen et al.\ 2005)
and the low-luminosity AGN NGC\,3998 (Sturm et al.\ 2005)
peak at a much longer wavelength ($\simali$11$\mum$),
inconsistent with ``standard'' silicate ISM dust
(which peaks at $\simali$9.7$\mum$).
The 9.7$\mum$ feature of NGC\,3998 is also much broader
than that of the Galactic ISM (Sturm et al.\ 2005).
The deviations of the silicate emission profiles of
type 1 AGNs from that of the Galactic ISM dust 
may indicate differences in the dust composition, 
grain size distribution, or radiative transfer effects 
(Sturm et al.\ 2005, Levenson et al.\ 2007).
The red tail of the 18$\mum$ silicate feature of NGC\,3998 
is significantly weaker than that of the bright quasars
(Sturm et al.\ 2005), suggesting that there may exist 
significant environmental variations.
Finally, it is worth noting that the 9.7$\mum$
silicate feature of Mkn 231, a peculiar
type 1 Seyfert galaxy, is also seen in {\it absorption}
peaking at $\simali$10.5$\mum$ (Roche et al.\ 1983).

\vspace{-3mm}
\subsection{The 3.3, 6.2, 7.7, 8.6 and 11.3$\mum$ PAH
            Emission Features}
\vspace{-1mm}
The distinctive set of ``Unidentified Infrared''
(UIR) emission features at 3.3, 6.2, 7.7, 8.6, and 11.3$\mum$,
now generally identified as the vibrational modes of PAH molecules 
(L\'{e}ger \& Puget 1984; Allamandola et al.\ 1985), 
are seen in a wide variety of Galactic and extragalactic regions
(see Draine \& Li 2007). 
In the Milky Way diffuse interstellar medium (ISM), PAHs, 
containing $\simali$45$\ppm$ (parts per million, relative to H) C,
account for $\simali$20\% of the total power emitted 
by interstellar dust (Li \& Draine 2001b).
The {\it ISO} (Infrared Space Observatories) 
and {\it Spitzer} imaging and spectroscopy 
have revealed that PAHs are also a ubiquitous feature of 
external galaxies. 
Recent discoveries include 
the detection of PAH emission in a wide range of systems:
distant Luminous Infrared Galaxies (LIRGs) 
with redshift $z$ ranging from 0.1 to 1.2 
(Elbaz et al.\ 2005),
distant Ultraluminous Infrared Galaxies 
(ULIRGs) with redshift $z\sim$\,2 (Yan et al.\ 2005),
distant luminous submillimeter galaxies at 
redshift $z\sim$\,2.8 (Lutz et al.\ 2005),
elliptical galaxies with a hostile environment
(containing hot gas of temperature $\sim$\,10$^7\K$) 
where PAHs can be easily destroyed 
through sputtering by plasma ions
(Kaneda et al.\ 2005),
faint tidal dwarf galaxies with metallicity 
$\sim Z_\odot/3$ (Higdon et al.\ 2006),
and galaxy halos (Irwin \& Madden 2006,
Engelbracht et al.\ 2006).

However, the PAH features are absent in AGNs,
as first noticed by Roche et al.\ (1991).
This is commonly interpreted 
as the destruction of PAHs 
by extreme UV and soft X-ray photons in AGNs 
(Roche et al.\ 1991; Voit 1991, 1992; 
Siebenmorgen et al.\ 2004).
Genzel et al.\ (1998) proposed to 
use the line-to-continuum ratio of
the 7.7$\mum$ PAH feature as a discriminator
between starburst and AGN activity in ULIRGs
(i.e. whether the dominant luminosity source of ULIRGs 
is an AGN or a starburst).
We should note that the PAH emission features
are detected in some Seyfert 2 galaxies, 
but they are from the circumnuclear star-forming 
regions, not from the AGNs
(e.g. see Le Floc'h et al.\ 2001, Siebenmorgen et al.\ 2004).

\vspace{-2.5mm}
\subsection{The Ice Absorption Features}
\vspace{-1mm}
Grains in dark molecular clouds 
(usually with $A_V$$>$3$\magni$) obtain ice mantles
consisting of H$_2$O, NH$_3$, CO,
CH$_3$OH, CO$_2$, CH$_4$, H$_2$CO 
and other molecules (with H$_2$O as the dominant species),
as revealed by the detection of various ice absorption 
features (e.g., H$_2$O: 3.1, 6.0$\um$; CO: 4.67$\um$; 
CO$_2$: 4.27, 15.2$\um$; CH$_3$OH: 3.54, 9.75$\um$; 
NH$_3$: 2.97$\um$; CH$_4$: 7.68$\um$;
H$_2$CO: 5.81$\um$; XCN$^{-}$: 4.62$\um$).
The ice absorption features are also seen
in most ULIRGs (e.g. see Spoon et al.\ 2002),
indicating the presence of a large quantity 
of molecular material in ULIRGs.
However, the ice absorption features 
are not expected in AGNs due to the high
dust temperatures (because of the immense 
bolometric luminosity emitted from the AGN)
-- the dust in the torus,
even at a distance of $\simali$100\,pc, 
is too warm ($>$\,100\,K)
for ice mantles to survive.

\vspace{-5mm}
\section{IR Emission Modeling\,--\,Inferring Dust Size and Torus Geometry?}
\vspace{-3mm}
To constrain the dust size distribution and 
the size and geometry of the dust torus,
various models have been proposed to explain 
the observed IR emission spectral energy distribution (SED)
of AGNs, radiated by the circumnuclear dust heated
by the AGN illumination.
These models assume a wide range of torus geometries:
uniform density annular (cylindrical) rings
of a few pc with an extremely large optical depth
$\tau_{\rm UV}$\,$>$\,1000 (Pier \& Krolik 1992, 1993),
optically thick plane parallel slabs
of a few thousand pc (Laor \& Draine 1993),
extended tori of hundreds of pc 
(Granato \& Danese 1994, Granato et al.\ 1997), 
geometrically thin, optically thick spherical shells
(Rowan-Robinson 1995),
tapered disks (Efstathiou \& Rowan-Robinson 1995,
Stenholm 1995),
optically thick, flared disks (Manske et al.\ 1998),
clumpy tori (Nenkova et al.\ 2002),
and other more complicated torus geometries
(van Bemmel \& Dullemond 2003, Schartmann et al.\ 2005).
In order to suppress the 9.7$\mum$ silicate emission feature
(which was not detected until very recently 
by {\it Spitzer}; see \S3.3),
some models hypothesized that the dust in AGNs must be large
($a$\,$<$\,10$\mum$) or small silicate grains must be depleted
(e.g. see Laor \& Draine 1993, Granato \& Danese 1994).
Some models ascribed the suppression of the 9.7$\mum$ silicate 
emission feature to clumpiness (Nenkova et al.\ 2002)
or the strong anisotropy of the source radiation
(Manske et al.\ 1998). 
Apparently, more modeling efforts are required to
account for the very recent detection of the 9.7$\mum$
and 18$\mum$ silicate emission features in type 1 AGNs
and the recent high resolution IR imaging observations which
seem to show that the torus size is no more than 
a few parsecs (see Elitzur 2006 and references therein).
It is well known that the SED modeling alone does not
uniquely determine the dust size distribution 
and the dust spatial distribution.

\acknowledgements I thank L.C. Ho and J.M. Wang  
for inviting me to attend this stimulating conference. 
I also thank B. Cznery, C.M. Gaskell, S.L. Liang,
R. Maiolino, and C. Willott for their comments and/or 
help in preparing for this article.
Partial support by NASA/Spitzer theory programs and
the University of Missouri Research Board 
is gratefully acknowledged.

\end{document}